\newcommand{\Xmax}{X_{\rm max}}
\newcommand{\avgXmax}{\langle{X_{\rm max}}\rangle}

\documentclass[preprint,12pt]{elsarticle}

\usepackage{units}

\usepackage{amssymb}

\makeatletter
\def\ps@pprintTitle{%
 \let\@oddhead\@empty
 \let\@evenhead\@empty
 \def\@oddfoot{\reset@font\hfil\thepage\hfil}
 \let\@evenfoot\@oddfoot
}

\journal{ }

\begin{document}

\begin{frontmatter}



\title{The non-linearity between $\langle\ln A\rangle$ and
  $\langle\Xmax\rangle$ induced by the acceptance of fluorescence
  telescopes}

\author[KIT]{R. Ulrich}
\author[LIP]{L. Cazon\corref{cor1}}
\ead{cazon@lip.pt}
\cortext[cor1]{Corresponding author  {\it Tel} +351 217973880 {\it Fax} +351 217934631}
\address[LIP]{LIP, Av. Elias Garcia, 14-1, 1000-149 Lisboa, Portugal}
\address[KIT]{Institut f\"ur Kernphysik, Postfach 3640, 76021 Karlsruhe, Germany}







\begin{abstract}

The measurement of the average depth of the shower maximum is the most commonly used observable for the possible inference of the primary cosmic-ray mass composition. Currently, different experimental Collaborations process and present their data not in the same way, leading to problems in the comparability and interpretation of the results. 
Whereas  $\langle\Xmax\rangle$ is expected to be proportional to $\langle \ln A \rangle$ in ideal conditions, we demonstrate that
the finite field-of-view of fluorescence telescopes plus the attenuation in the
atmosphere introduce a non-linearity into this relation, which is specific for each particular detector setup.

\end{abstract}


\end{frontmatter}


\section{Introduction}
\label{s:int}

Ultra-high energy cosmic-ray (UHECR) particles are the highest energy
particles observed by humankind, well beyond what is accessible at the
Large Hadron Collider. Current experiments \cite{Auger,HiRes,TA} are
recording large amounts of high quality data. Recent observations do
not draw a simple nor consistent picture of the nature of UHECR
particles. There are hints of anisotropy in the arrival
directions~\cite{AugerAnisotropy}, favouring the presence of light primary
particles. Using the current high energy hadronic interaction models, typical air-shower observables do not generally
support the light component hypothesis: the Pierre Auger Collaboration observes the depth of the electromagnetic
shower maximum, muon production maximum, rise-time of the shower front,
etc.\ that are all pointing in the direction of a primary composition
dominated by heavy particles at ultra-high
energies~\cite{AugerER,Auger_ICRC}. On the other hand, the HiRes
Collaboration claims that their measurement of the depth of the shower maximum is compatible with
protons up to the highest energies~\cite{Abbasi:2009nf}, which is currently supported by
the first TA data~\cite{FirstTAdata,TAER}. 
It is important to notice that whereas Auger publishes $\avgXmax$ in the atmosphere by using a fiducial volume selection to minimize acceptance biases, HiRes and TA presently do not apply such corrections and present $\avgXmax$ at the detector level.
Also the possible difference of the energy scale between Auger and HiRes/TA ($\approx$25\,\%)~\cite{Abraham:2010mj} contribute to the difficulties to explain the data~\cite{Engel:2011zz,OlintoTAUP2011}.

The current situation could point to exciting physics aspects that are about to emerge with higher accumulated event statistics, as for instance, systematic differences of the cosmic-ray flux on the northern and southern hemisphere.  The Pierre Auger Observatory observes the southern, while HiRes/TA the norther sky, with very
different astrophysical objects in their direct field-of-view. 
It is clear, that the understanding of the primary mass composition of
UHECR is one of the major steps towards the final solution of the UHECR
puzzle. Depending on a reliable measurement of the mass composition
very different scenarios of the nature of UHECR will finally emerge.

 In this article we discuss  the concept of \emph{observed} versus \emph{true} $\Xmax$-distributions to finally demonstrate that the  \emph{observed} average of depth of shower maxima, $\langle\Xmax\rangle$, can result in a non-linear relation with  the average logarithm of the mass number, $\langle \ln A \rangle$, of the UHECR. These effects are very specific for a
given experimental setup and have to be accounted for in order to
allow a comparison or interpretation of these data.

\section{Extensive Air-Showers and Fluorescence Telescope}

Fluorescence telescopes measure the ultraviolet light, which is proportional to the energy deposited by the passage of the charged particles of the air-shower cascade through the atmosphere. This makes possible to reconstruct the longitudinal profile of the electromagnetic shower~\cite{Unger}. 
 The atmospheric depth at which the energy deposit is maximal is the shower maximum, $\Xmax$, and is related to the nature of the primary particle.
At the same primary energy, primary nuclei with mass $A$ produce air-showers with shower maxima at different atmospheric depths. This can be approximated as~\cite{Matthews} 
\begin{eqnarray}
  \nonumber
  \Xmax\approx \lambda_{\rm int} + \ln2\, X_0 \ln \frac{E}{A},
\end{eqnarray}
where $\lambda_{\rm int}$ is the cross-section of cosmic-ray primaries with air and $X_0\sim\unit[37]{g/cm^2}$ is the electromagnetic radiation length in air. 
The structure of this equation, $\Xmax=a + b \ln E/A$, holds also for more general considerations~\cite{AlvarezMuniz:2002ne}. It has been typically used to compute the average logarithmic mass from the data of the average depth of the shower maximum
\begin{eqnarray}
  \nonumber
  \langle\ln A\rangle = \frac{1}{b} \left[ a + b\ln E - \langle\Xmax\rangle \right] = \frac{1}{b} \left[ \langle\Xmax\rangle_{p}-\langle\Xmax\rangle \right] .
  \label{eq:lnAXmax}
\end{eqnarray}
Thus, there is a linear dependence of the measured average shower maximum 
from the average mass of the cosmic-ray primary particles.

Instead of calculating $\langle\ln A\rangle$ one can also compute the component fractions in a model with two primary cosmic-ray species. This is typically done for the proton fraction $f$ under the assumption of a simple proton/iron mixture. 
For this case the relation of $\langle\Xmax\rangle$ is
\begin{equation}
  \langle\Xmax\rangle=f \langle\Xmax\rangle_p+(1-f)\langle\Xmax\rangle_{\rm Fe} ,
  \label{eq:linear}
\end{equation}
which is linear in $f$. 

Thus, the distinction between an admixture of different primaries or a pure primary of an intermediate mass, are indistinguishable on the base of a $\langle\Xmax\rangle$ value alone. The resulting equivalence is
\begin{eqnarray}
  \nonumber
  \langle \ln A \rangle= \frac{1}{b} \left[\langle\Xmax\rangle_p-\langle\Xmax\rangle_{\rm Fe}\right] (1-f).
\end{eqnarray}

\section{The effect of the telescope acceptance}

The objective of this paper is not to reproduce any particular detector configuration, but to demonstrate that the relation between $\langle \Xmax \rangle$ and $\langle \ln A \rangle$ depends critically on the detector acceptance, which varies with the primary energy. Thus, the telescope acceptance has an impact on the interpretation of $\langle\Xmax\rangle$ data, and in general any other momentum of the $\Xmax$-distribution~(see also~\cite{unger2}).

Fluorescence telescopes have a limited viewing angle defined by the optics. Showers falling outside the field-of-view cannot be detected. But even for observed showers the shower maximum can be outside the field-of-view of the telescope and can then not be reconstructed reliably.
\begin{figure}[!]
  \centering
  \includegraphics[width=.45\textwidth]{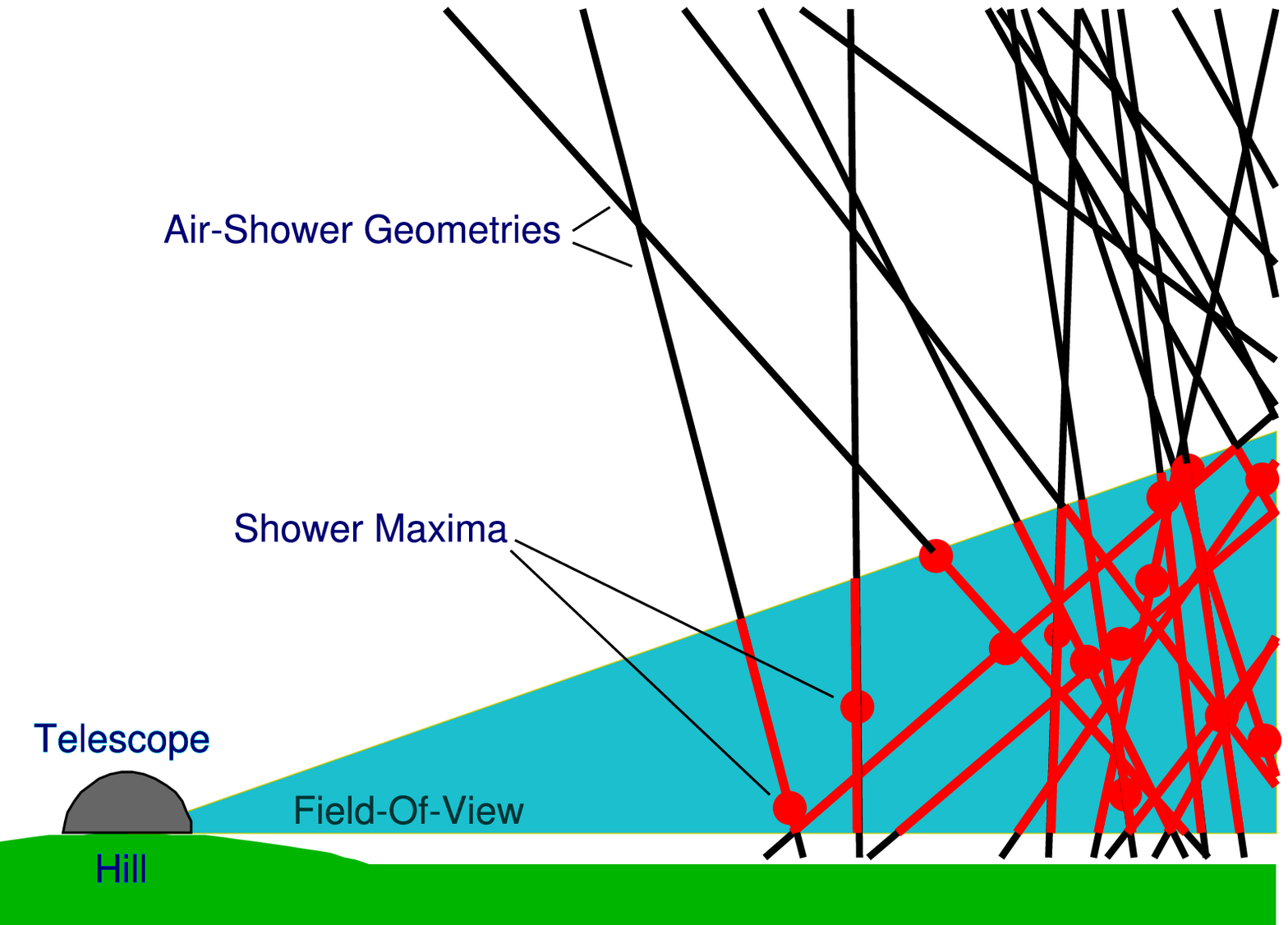}~\hfill
  \includegraphics[width=.5\textwidth]{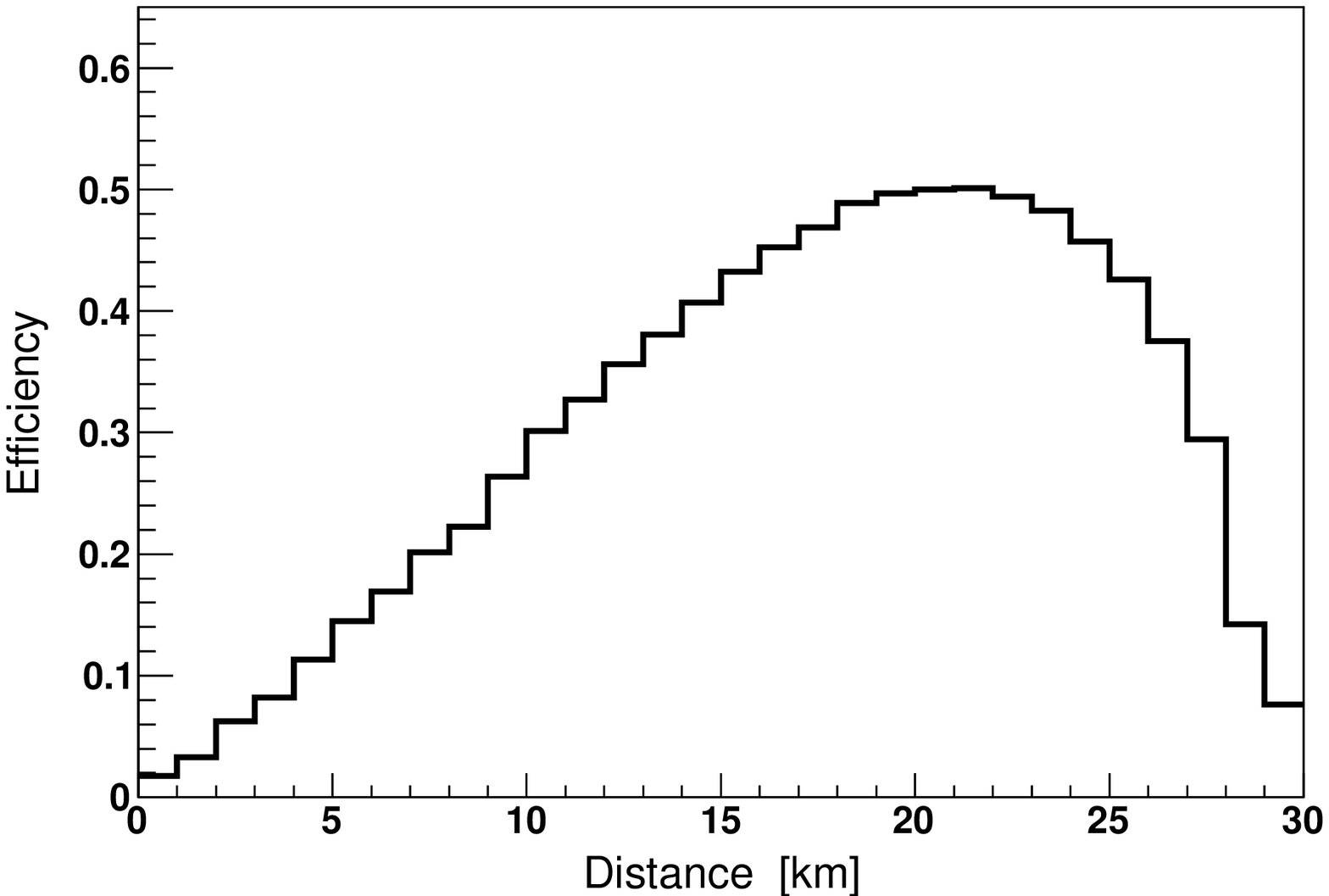}\\
  \caption{Left panel: Geometry of our simulation study. Right panel: Resulting efficiency, $N_{\rm obs}/N_{\rm gen}$, depending on the distance of air-shower cores from the telescope. \label{fig:fov}}
\end{figure}
The angular field-of-view defines a volume in the atmosphere where 
showers can effectively be fully reconstructed. One of the most important 
criteria of this is that the shower maximum is located within the field-of-view. 
This volume is depicted in Fig.~\ref{fig:fov}~(left).  When showers  fall very close to the telescope, the transverse area of the field-of-view is small, meaning that many shallow and/or deep showers are not fully reconstructed. This effect is naturally most severe in the energy range close to the lower detection threshold for air-showers where the distance to the events is limited by the small amount of generated fluorescence light. For air-showers detected at larger distance to the telescope the geometrical field-of-view cone is much larger. However, fluorescence light can travel only a limited distance in the atmosphere before being absorbed. At some point, light emitted at the shower axis is attenuated too much and cannot be observed any more. This is responsible for the characteristic rapid drop of  the efficiency as shown in Fig.~\ref{fig:fov}~(right) beyond 20\,km.
The Rayleigh absorption rate is proportional to the density of the atmosphere, 
and thus depends on changing air pressure as well as on the height in the atmosphere 
at which a particular shower is developing. In this work we do not discuss 
aerosol related absorption. It is much smaller than Rayleigh absorption and 
to include it here does not help the argument, but just adds additional complexity.

By means of a toy Monte Carlo we reproduce all parameters relevant for the modelling of air-shower observation with fluorescence telescopes at an arbitrary primary energy. Our simulations focus on the geometric and atmospheric effects and simplifies the situation as much as possible. We do not attempt to reproduce a specific detector setup, but only want to show the inevitable impact on the interpretation of the observed data.

The geometry of our simulation is shown in Fig.~\ref{fig:fov}. The
vertical depth of the telescope is at $\unit[\approx880]{g/cm^2}$,
which corresponds to h=1.5\,km, as explained below, and the
opening angle of the field-of-view is 20$^\circ$. We consider air-showers of fixed primary
energy, with random arrival directions, sampled from ${\rm d}N/{\rm
  d}\cos\theta=$const up to 60$^\circ$ zenith angle, as well as distances $l$
from the telescope, sampled from ${\rm d}N/{\rm
  d}l\propto{l}$. The requirement of a particular shower to be
observed is that at least $N_{\rm det}=100$ photons reach the
telescope from the location of the shower maximum on the shower
axis. The number of photons is computed as
\begin{eqnarray}
  N_{\rm det} = N_{\rm ph} A_{\rm dia} / r^2 \exp\left(-\frac{t}{\lambda_{\rm abs}}\right),
  \label{eq:ndet}
\end{eqnarray}
where $N_{\rm ph}$ is the number of fluorescence photons emitted at
the shower maximum, $A_{\rm dia}=10\,$m$^2$ is the aperture of the
telescope, $r$ is the geometric distance, $t=\int\rho(z){\rm
  d}\vec{r}$ the integrated depth distance from the telescope to the
location of $X_{\rm max}$, and $\lambda_{\rm abs}=\unit[1000]{g/cm^2}$
is the photon absorption length. For air showers of primary energy
$E_0$ and a fluorescence yield of $Y_{\rm fluo}=5/$MeV the number of photons
emitted at $X_{\rm max}$ is 
\begin{equation}
  N_{\rm ph}=\left(\frac{{\rm d}E}{{\rm d}X}\right)_{\rm max}\; Y_{\rm fluo} \; \rho \; {\rm c} \; \Delta t\,,
  \label{eq:nph}
\end{equation}
where $\Delta t=100\,$ns is the telescope sampling time and $({\rm
  d}E/{\rm d}X)_{\rm max}$ is the energy deposit at the shower
maximum, which is $10^{-11.77 + \lg(E_0/{\rm eV})}\,$GeV\,cm$^2$/g in
very good approximation\footnote{This is obtained with the SIBYLL~\cite{Ahn:2009wx}
  interaction model.}. For air showers with $E_0=10^{18.5}\,$eV this
yields $N_{\rm ph}\approx2\cdot10^{10}$, which is the default value
throughout this paper if not stated otherwise.  The atmospheric
density profile $\rho(z)$ is exponential with a scale height of
\unit[8]{km} and a pressure at the height above sea level,
$h=\unit[1.5]{km}$, of the telescope of $\unit[\approx 880]{g/cm^2}$.

\begin{figure}[bth!]
  \begin{center}
    \includegraphics[width=.8\textwidth]{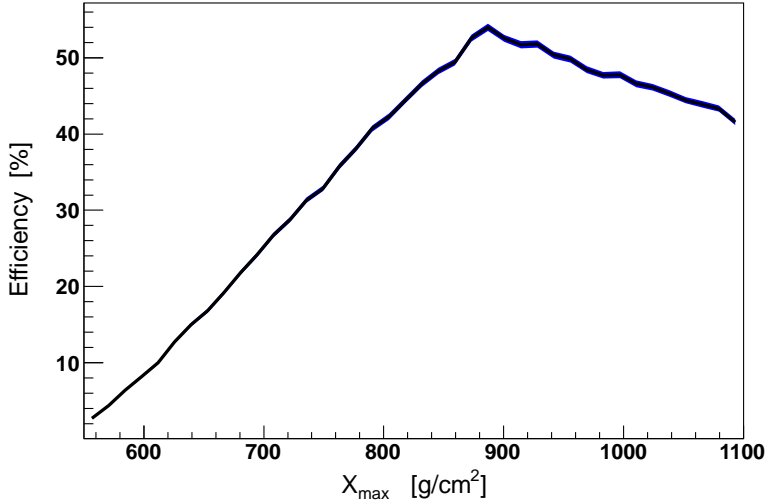}
    \caption{Efficiency $\epsilon(\Xmax)$ of the $\Xmax$ reconstruction as a function of
      $\Xmax$  for a particular detector, as
      calculated by our toy Monte Carlo.\label{fig:eff}}
  \end{center}
\end{figure}

When we generate events with a flat distribution of $X_{\rm max}$, we
find that air-showers with different $X_{\rm max}$ are observed with
different efficiency $\epsilon(X_{\rm max})=N_{\rm obs}/N_{\rm
  gen}$. In Fig.~\ref{fig:eff} we show this efficiency as a function
of $X_{\rm max}$. The peak of this distribution is related to the
\emph{average atmospheric slant depth} in the volume observed by the
telescope. Every telescope setup has limits both for very small as
well as very large values of $X_{\rm max}$ (see
e.g.\ \cite{flyseye}). While large integrated atmospheric depths can
in principle always be achieved by very inclined geometries, there is
a strict bound on the minimal observed depth even for vertical events,
which is related to the maximum possible observation distance and the
upper elevation boundary of the field-of-view. This effect becomes
more relevant for lower energy air-showers, since here the maximum
observation distance is smaller.

\begin{figure}[tb!]
  \begin{center}
    \includegraphics[width=.7\textwidth]{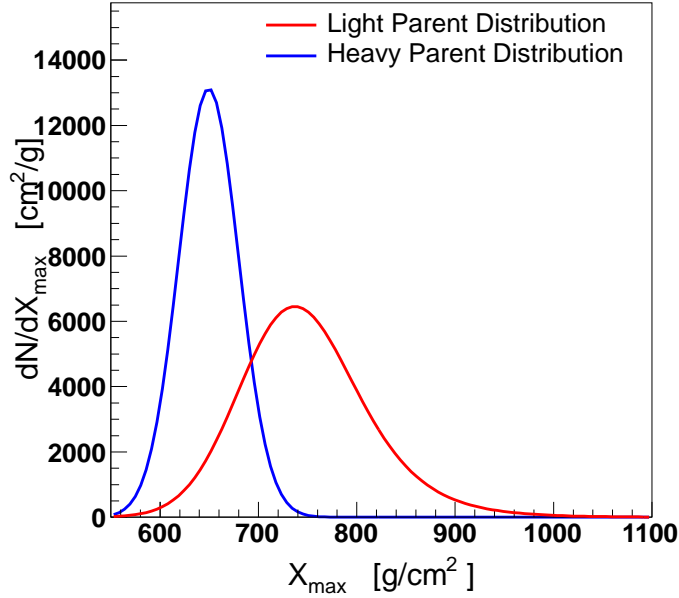}
    \caption{The two input distributions corresponding to light and
      heavy primary cosmic-ray particles in our study. The
      distributions are Exponential convoluted with a Gaussian.  The
      used parameters of Eq.~(\ref{eq:modelxmax}) are
      $\mu=\unit[815]{g/cm^2}$, $\sigma=\unit[50]{g/cm^2}$ and
      $\tau=\unit[45]{g/cm^2}$ for the light component as well as
      $\mu=\unit[1005]{g/cm^2}$, $\sigma=\unit[30]{g/cm^2}$ and
      $\tau=\unit[5]{g/cm^2}$ for the heavy component.
      \label{fig:input}}
  \end{center}
\end{figure}

Given the average efficiency $\epsilon(X_{\rm max})$ of a telescope
setup, the measured distribution of $X_{\rm max}$ is related to the parent distribution via
\begin{eqnarray}
  \nonumber \left(\frac{{\rm d}N}{{\rm d}X_{\rm max}}\right)_{\rm
    measured} = \epsilon(X_{\rm max}) \left(\frac{{\rm d}N}{{\rm
      d}X_{\rm max}}\right)_{\rm true}.
\end{eqnarray}
Small changes in the telescope setup can yield very different
$\epsilon(X_{\rm max})$.  If also energy, $X_{\rm max}$ resolution,
and possible $X_{\rm max}$ reconstruction biases effects were included
the relation between the measured and parent distribution would become
more complex~\cite{resolution, vitor}, but this does not help the
clarity of our argument.

In the following we demonstrate how the telescope acceptance,
$\epsilon(X_{\rm max})$, affects in particular also \emph{simple}
analyses as for example related to $\langle X_{\rm max}\rangle$. For
this purpose we generate input $X_{\rm max}$ distributions derived
from an Exponential convoluted with a Gaussian function
\begin{eqnarray}
  \frac{{\rm d}N}{{\rm d}X_{\rm max}} = \frac{N_{\rm evt}}{2\tau} \;\;\,{\rm e}^{\sigma^2/(2\tau^2)}\;\; {\rm e}^{(\mu-X_{\rm max})/\tau} \;\;  {\rm erfc}\left(\frac{\mu-X_{\rm max}-\sigma^2/\tau}{\sqrt{2}\sigma}\right).
  \label{eq:modelxmax}
\end{eqnarray}
In total this function has three shape parameters: the mean and width
of the Gaussian, $\mu$ and $\sigma$, and the exponential slope,
$\tau$. We generate two distributions (c.f.\ Fig.~\ref{fig:input}),
the first one corresponding to \emph{light} cosmic-ray primaries has
an average of $\unit[750]{g/cm^2}$ and an RMS of $\unit[66.5]{g/cm^2}$,
the second one corresponds to \emph{heavy} cosmic-ray primaries has an
average of $\unit[650]{g/cm^2}$ and an RMS of $\unit[30]{g/cm^2}$. 
These values are chosen since
they correspond roughly to typical values for proton or iron induced
air-showers respectively. The average efficiencies
\begin{eqnarray}
  \nonumber
  \epsilon_{A}=\frac{\int\limits_0^{\infty} \left(\frac{{\rm d}N}{{\rm d}\Xmax}\right)_{\rm measured} {\rm d}\Xmax}{\int\limits_0^{\infty} \left(\frac{{\rm d}N}{{\rm d}\Xmax}\right)_{\rm true} {\rm d}\Xmax} = \int\limits_0^\infty \epsilon(\Xmax) {\rm d}\Xmax
\end{eqnarray}
for the observation of $X_{\rm max}$ with our telescope setup for
these two distributions are $\epsilon_{\rm light}=33.1\,\%$ and
$\epsilon_{\rm heavy}=16.7\,\%$. The average $X_{\rm max}$ observed by
the telescope setup are $\langle\Xmax\rangle^\prime_{\rm
  light}=\unit[770]{g/cm^2}$ for the light and
$\langle\Xmax\rangle^\prime_{\rm heavy}=\unit[660]{g/cm^2}$ for the
heavy component, which is $\unit[10-20]{g/cm^2}$ biased with respect
to the input distributions.
\begin{figure}[bt!]
  \begin{center}
    \includegraphics[width=.9\linewidth]{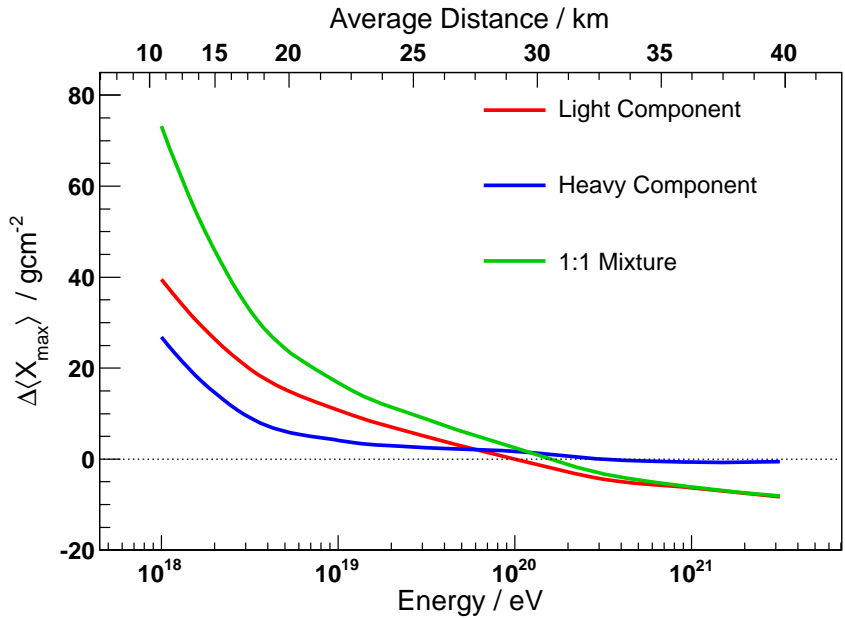}
    \caption{Dependence of the bias $\Delta\langle X_{\rm max}\rangle$
      from the energy and thus the distance of air-showers from the
      telescope. The upper axis indicates the resulting average
      distance of showers from the telescope depending on the
      primary energy.\label{fig:energy}}
  \end{center}
\end{figure}
\begin{figure}[bt!]
  \begin{center}
    \includegraphics[width=.9\linewidth]{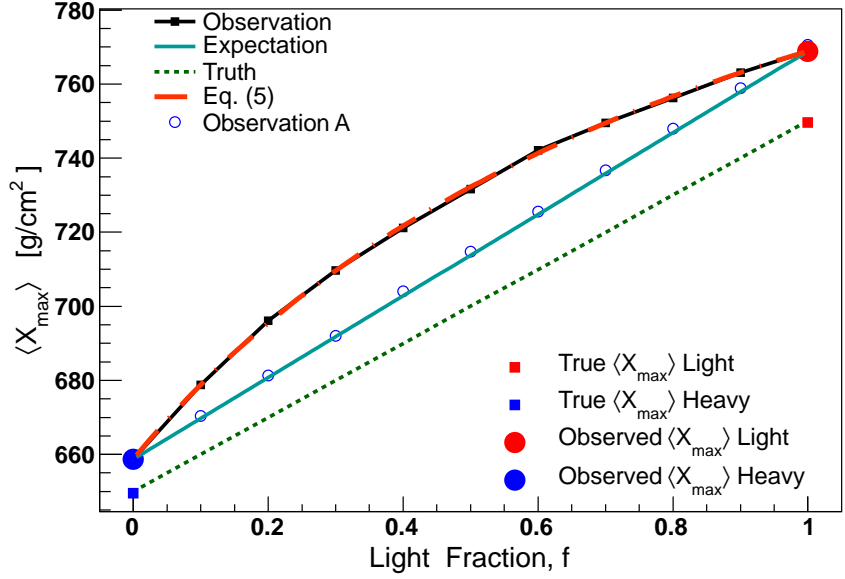}
    \caption{The average $\Xmax$ as a function of the light cosmic-ray
      fraction, $f$. The \emph{observed} $\avgXmax$ values for pure
      light and pure heavy compositions are shifted with respect to
      the \emph{true} $\avgXmax$. The relation between $\avgXmax$ and
      $f$, and therefore with $\langle \ln A \rangle$ by
      Eq. \ref{eq:lnAXmax}, is linear for the true $\Xmax$
      distributions, whereas it is non-linear for the case of the
      observed $\avgXmax$. The curve labeled ``Observation A''
      indicates the effect on a pure composition of intermediate mass.}
    \label{f:fXmax}
  \end{center}
\end{figure}
The bias $\Delta\langle X_{\rm max}\rangle$, defined as the difference
between the observed and true $\langle X_{\rm max}\rangle$ value,
changes with distance to the telescope and thus also with the primary
energy of the cosmic-ray particles according to Eqs.~(\ref{eq:ndet}) and (\ref{eq:nph}). In Fig.~\ref{fig:energy} this is
shown for the cases of pure light, pure heavy and an equal mixture of
light and heavy primaries.  For this study we consider a change of
$\langle X_{\rm max}\rangle$ with energy of ${\rm d}\langle X_{\rm
  max}\rangle/{\rm d}\lg{E}=\unit[60]{g/cm^2}$ with respect to the
default values used otherwise in this paper.  It is interesting to
note that the resulting bias depends on the underlying mass
composition.  In general, for smaller energies the showers are located
closer to the telescope and the effect becomes stronger. For example,
the low energy fluorescence telescope extensions HEAT~\cite{heat} at
the Pierre Auger Observatory and TALE~\cite{tale} at the Telescope
Array are important to limit these biases by providing a much wider
field-of-view for showers at close distances.  Furthermore, the bias
can be positive as well as negative, depending on the overlap of the
true $X_{\rm max}$ distribution with the telescope efficiency
function, c.f.\ Fig.~\ref{fig:eff}. However, the impact will
qualitatively be always as shown in Fig.~\ref{fig:energy}: at short
distances the field-of-view cuts shallow showers and thus $\langle
X_{\rm max}\rangle$ is overestimated, at larger distances this effects
becomes smaller and eventually might even reverse leading to an
underestimation of the true $\langle X_{\rm max}\rangle$ value.
It is important to realize that the primary effect is due to
a different distance of showers, and that any parameter that affects
the typical observation distance will have a similar impact. Such
parameters are typically related to the detector setup, for example
the diaphragm opening $A_{\rm dia}$, the sampling time $\Delta t$, but
also the atmospheric density profile and optical absorption
characteristics.

\begin{figure}[bt!]
  \begin{center}
    \includegraphics[width=.9\linewidth]{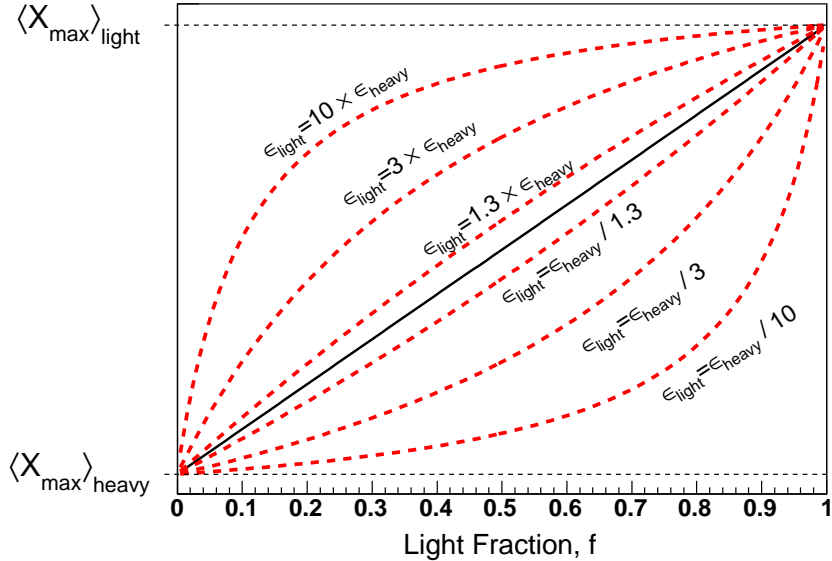}
    \caption{Impact of different values of $r=\epsilon_{\rm light}/\epsilon_{\rm
        heavy}$ on the non-linearity.}
    \label{fig:scan}
  \end{center}
\end{figure}

If we consider a mixture of light and heavy particles at fixed primary
energy with a given fraction $f$ of light and $1-f$ of heavy primary
particles, the average observed shower maximum becomes
\begin{equation}
  \langle\Xmax\rangle^\prime=\frac{f
    r \langle\Xmax\rangle^\prime_{\rm light} +(1-f) \langle\Xmax\rangle^\prime_{\rm heavy} }{fr + 1-f} 
  \label{eq:nonlinear}
\end{equation}
where $r=\epsilon_{\rm light}/\epsilon_{\rm heavy}$. This
relation is non-linear in $f$ and thus not in $\ln A$. In
Figure~\ref{f:fXmax} we show the results of this study. The true as
well as the observed $\avgXmax$ are shown, together with the linear
interpolation from Eq.~(\ref{eq:linear}) and the non-linear version
Eq.~(\ref{eq:nonlinear}). The latter describes perfectly well the
results of our Monte Carlo study.

We also study how a different detector setup affects our result.  In general, by
changing parameters of the telescope detector setup, very different
values of $r$ can be obtained. In Fig.~\ref{fig:scan} we show
how the predictions of Eq.~(\ref{eq:nonlinear}) change for a wide range of different
values of $r$.  It summarizes the effect described in this
paper, and the miss-interpretation it can induce on the inference of
the primary mass composition from $\langle \Xmax \rangle$. For
instance, $r=3$ would imply that a real fraction 50\,\% proton
and 50\,\% iron, that is f=0.5 and $\langle\ln A\rangle=2$, would be
interpreted as 75\,\% proton and 25\,\% iron, $\langle\ln A\rangle=1$, if
just the linear relation Eq.~(\ref{eq:linear}) is assumed.

\section{Summary}
The data collected by a fluorescence telescope are affected by
acceptance effects, which can have an important impact on the detailed
interpretation of the data.  This is problematic for researchers
not from within a particular experimental collaboration who, for
example, have no access to a detailed Monte Carlo simulation
of the detector setup.

The average efficiency $\epsilon(X_{\rm max})$, which describes the
response to a flat distribution in $X_{\rm max}$, is one of the crucial
properties of a fluorescence telescope.  However, in addition to the
acceptance effects also the detector resolution and $X_{\rm max}$
reconstruction biases have to be known for a
full data analysis.

We have demonstrated that the bias in the average shower maximum
introduced by the acceptance of fluorescence telescopes induces a non
linearity between $\avgXmax$ and $\langle \ln A \rangle$.

\section{Acknowledgments}

We would like to thank A. Olinto for the initial discussions that
inspired this work during the TAUP2011 meeting at Munich. J.~Bellido, D.~Harari, M.~Bueno, M.~Unger,
R.~Concei\c{c}\~{a}o and M.~Pimenta for careful reading of this manuscript and their
comments. Finally we thank our colleagues from the Pierre Auger
Collaboration for helping us to get the necessary insight on the
fluorescence detection techniques. L.C.\ wants to thank fundings by
Funda\c{c}\~{a}o para a Ci\^{e}ncia e Tecnologia (CERN/FP11633/2010),
and fundings of MCTES through POPH-QREN Tipologia 4.2, Portugal, and
European Social Fund.

\bibliographystyle{elsarticle-num}

\begin{thebibliography}{99}

 \bibitem{Auger} 
  J.~Abraham {\it et al.}  [Pierre Auger Collaboration],
  Nucl.\ Instrum.\ Meth.\  A {\bf 620}, 227 (2010).

 \bibitem{HiRes} 
  R.U.~Abbasi {\it et al.},
  Astropart.\ Phys.\  {\bf 32}, 53 (2009).


 \bibitem{TA} 
  M.~Takeda  {\it et al.} [TA Collaboration],
  Mod.\ Phys.\ Lett.\  A {\bf 23}, 1301 (2008).


 \bibitem{AugerAnisotropy}
  P.~Abreu {\it et al.} [Pierre Auger Collaboration],
  JCAP {\bf 1106}, 022 (2011).




 \bibitem{AugerER} 
  J.~Abraham {\it et al.}  [Pierre Auger Collaboration],
  Phys.\ Rev.\ Lett.\  {\bf 104}, 091101 (2010).

 \bibitem{Auger_ICRC} 
  P.~Facal, D.~Garcia-Pinto, D.~Garcia-Gamez for The Pierre Auger Collaboration, 32nd International Cosmic Ray Conference, Beijing, China, August 2011, 
  arXiv:1107.4804.

\bibitem{Abbasi:2009nf}
  R.U.\ Abbasi \textit{et al.} [HiRes Collaboration],
      Phys.\ Rev.\ Lett. \textbf{104}, 161101 (2010).


\bibitem{FirstTAdata} 
  J.N.~Matthews [TA Collaboration],
  Nucl.\ Phys.\ Proc.\ Suppl.\  {\bf 212-213}, 79 (2011). 

 \bibitem{TAER} D.~Ikeda to appear in the proceedings of the 12th internation conference of Topics in Astroparticle and Underground Physics, TAUP2011, Munich, 2011.

\bibitem{Abraham:2010mj}
  J.\ Abraham \textit{at al.} [Pierre Auger Collaboration], 
  Phys. Lett. B \textbf{685}, 239 (2010).


\bibitem{Engel:2011zz}
  R.~Engel [Pierre Auger Collaboration],
  AIP Conf.\ Proc.\  {\bf 1367}, 50 (2011).

 \bibitem{OlintoTAUP2011} A.~Olinto, to appear in the proceedings of the 12th international conference of Topics in Astroparticle and Underground Physics, TAUP2011, Munich (2011).


\bibitem{Unger} 
M. Unger, B.R. Dawson, R. Engel, F. Schussler, R. Ulrich,
Nucl.\ Instrum.\ Meth. A \textbf{588}, 433 (2008).

 \bibitem{Matthews}
  J.~Matthews,
  Astropart.\ Phys.\  {\bf 22}, 387 (2005).




\bibitem{AlvarezMuniz:2002ne}
J. Alvarez-Muniz, R. Engel, T. Gaisser, J. Ortiz, and
T. Stanev, Phys. Rev. D \textbf{66}, 033011 (2002).

\bibitem{unger2} M. Unger for the Pierre Auger Collaboration, 
  Nucl. Phys. B (Proc. Suppl.) \textbf{190}, 240 (2009).


\bibitem{Ahn:2009wx} E.\ Ahn \textit{et al.}, Phys.\ Rev.\ D \textbf{80}, 094003
  (2009).

\bibitem{flyseye}	
  G.L.\ Cassiday \textit{et al.},
  Astrophys. J. \textbf{356}, 669 (1990).


  \bibitem{resolution} J. Bellido for the Mass Composition Working Group, Talk at the \textit{International Symposium on Future Directions of UHECR Physics}, CERN, Geneva, 2012.

\bibitem{vitor}
 R. R. Prado and V. de Souza, \textit{Private comunication.}


 	
\bibitem{heat}
H.-J. Mathes for the Pierre Auger Collaboration, 32nd International Cosmic Ray Conference, Beijing, China, August 2011, arXiv:arXiv:1107.4804.


\bibitem{tale}
K. Martens for the Telescopes Array Collaboration, 
 Nucl.\ Phys.\ Proc.\ Suppl.\ \textbf{165}, 33 (2007).


 \end{thebibliography}



\end{document}